
%
%

\input harvmac.tex


\lref\berst {E. Bergshoeff, E. Sezgin and P. Townsend, Phys. Lett. {\bf B189}
 (1987)  75.}

\lref\achetw {A. Achucarro, J. Evans, P. Townsend and D. Wiltshire, Phys.
Lett. {\bf B198}  (1987)  441.}

\lref\duflfb {M. J. Duff and J. X. Lu, Nucl. Phys. {\bf B354}  (1991)  141.}

\lref\dabghr {A. Dabholkar, G. W. Gibbons, J. A. Harvey and F. Ruiz Ruiz,
Nucl. Phys. {\bf B340}  (1990)  33.}

\lref\dufhis {M. J. Duff, P. S. Howe, T. Inami and K. Stelle, Phys. Lett.
{\bf B191}  (1987)  70.}

\lref\dufs {M. J. Duff and K. Stelle, Phys. Lett. {\bf B253}  (1991)  113.}

\lref\calhswb {C. Callan, J. Harvey and A. Strominger, Nucl. Phys. {\bf B367}
 (1991) 60.}

\lref\duflthb {M. J. Duff and J. X. Lu, Phys. Lett. {\bf B273}  (1991)  409.}

\lref\hors {G. Horowitz and A. Strominger, Nucl. Phys. {\bf B360}  (1991)
197. }

\lref\duflpb {M. J. Duff and J. X. Lu, ``Black and super $p$-branes in diverse
dimensions", CERN-TH.6675/92 and CTP--TAMU--54/92  (1992).}

\lref\duflsds{M. J. Duff and J. X. Lu,``The self-dual Type IIB superstring"
 (unpublished).}

\lref\duflbscan{M. J. Duff and J. X. Lu, ``Type II $p$-branes: The Brane-Scan
Revisited", CERN-TH.6560/92 and CTP/TAMU-37/92 (to appear in Nucl. Phys.
{\bf B}).}

\lref\stra{J. Strathdee, Int. J. Mod. Phys. {\bf A2} (1987) 273.}

\lref\salsone{A. Salam and E. Sezgin, ``Supergravities in Diverse Dimensions",
North-Holland and World Scientific.}

\lref\luthesis{J. X. Lu, ``Supersymmetric Extended Objects", Texas A\&M,
Ph.D. Thesis (1992).}

\lref\salstwo {A. Salam and E. Sezgin,  Nucl. Phys. {\bf B258} (1985) 284.}

\lref\sen{A. Sen,  Phys. Lett. {\bf B274} (1992) 34.}

\lref\hass{S. F. Hassan and A. Sen,  Nucl. Phys. {\bf B375} (1992) 103.}

\lref\gids{S. B. Giddings and A. Strominger, Phys. Rev. Lett. {\bf 67} (1991)
2930.}

\def\dg{\hbox{$^\dagger$}}

\def\half{{1\over2}}

\Title{\vbox{\baselineskip12pt \hbox{CERN-TH.6877/93}}}
{\vbox{\centerline{\bf ADM Masses}\vskip2pt \centerline{\bf for Black Strings
and $p$-Branes}}}
\baselineskip=18pt

\centerline {J. X. Lu{\footnote\dg{Supported partially by
a World Laboratory Fellowship.}}}
\centerline{\it{CERN, Theory Division, CH-1211, Geneva 23, Switzerland}}
\centerline{jxlu@cernvm.cern.ch}
\bigskip
\centerline{\bf Abstract}

An ADM mass formula is derived for a wide class of black solutions with
certain spherical symmetry. By applying this formula, we calculate the ADM
masses for
recently
discovered black strings and $p$-branes in diverse dimensions. By this, the
Bogolmol'nyi equation can be shown to hold explicitly. A useful observation is
made for non-extremal black $p$-branes that only for $p = 0$, i.e. for a black
hole, can its ADM mass be read directly
from the asymptotic behaviour of the metric component $g_{00}$.

\bigskip
\vskip 1.5in
\bigskip
\noindent
CERN-TH.6877/93

\Date{April 1993}
\eject

\newsec{\bf Introduction}

     Black and extremal string and $p$-brane solutions have been
found recently by a number of authors \refs{\dabghr, \hors, \duflfb, \duflpb}.
As the metric $g_{MN}$ of a black configuration can be written as
$g_{MN} = \eta_{MN} + h_{MN}$, with $\eta_{MN}$ the flat Minkowski metric
and $h_{MN}$ not necessarily small everywhere, the total energy density
 has been given in
\dabghr, whose explicit form for extremal black strings has also been given.
The explicit form of the total energy density  for extremal black $p$-branes
was
given later, in \duflpb, where $p =1$, i.e. string as a special case. Then
the ADM mass for an extremal black $p$-brane follows the integration of the
corresponding explicit total energy density over the $(D - p -1)$-dimensional
transverse space.  As  is well known, the
usual black solution with certain spherical symmetry is most conveniently cast
 in
terms of some spherical coordinates. Black
strings and $p$-branes fall into this category, where the formula given in
\dabghr\ for the total energy density cannot be simply used. We will use the
standard definition of the gravitational energy-momentum
pseudo-tensor to derive the  ADM mass in the next section. In this short note,
we derive an ADM mass formula for
a class of black solutions with some spherical symmetry, then calculate the ADM
 masses
explicitly for the recently discovered black strings and $p$-branes. Finally,
we use the calculated ADM mass to show that the Bogolmol'nyi equation is
satisfied
for each of the discovered black strings and $p$-branes, which provides a way
to
justify the stability of the corresponding solutions. An observation is made of
when the ADM mass of a black configuration can be read directly from the
asymptotic behaviour of the $g_{00}$
component of the metric.

\newsec{\bf  ADM mass formula}

    In general relativity,  the local energy density of the gravitational
field cannot be defined uniquely, even in the weak-field limit. We will
adopt the standard definition of the gravitational energy-momentum
pseudo-tensor to find the ADM mass for a black hole, the ADM mass per unit
length
for a black string, and in general the ADM mass per unit volume for a
$p$-brane, in what
 follows. Let us write
$g_{MN} = g^{(0)}_{MN} + h_{MN}$, where $g^{(0)}_{MN}$ is the flat limit
of the corresponding space-time metric, for example, it could be Minkowski.
 In $D > p + 3$, $h_{MN}$ is
asymptotically zero but not necessarily small everywhere. In $D = p + 3$,
$h_{MN}$ is asymptotically logarithmically divergent. Following the discussion
for $p =1$, i.e. string in \dabghr, we pretend to take the  above definition
for
those cases, too. To  first order in $h_{MN}$, the Einstein equation looks
like
\eqn\emtensor{R^{(1)}_{MN} - \half g^{(0)}_{MN} R^{(1)} = \kappa^2
\Theta_{MN}.}
One can take this as the definition of the `` total" energy-momentum tensor,
to which the ADM mass per unit volume is defined as
\eqn\admass{ M_d = \int d^{D -d}y \Theta_{00},}
for a black $(d - 1)$-dimensional extended object. As
$g^{(0)}_{MN} = \eta_{MN}$, i.e. the  flat Minkowski,
the general $R^{(1)}_{MN}$ has been given in \dabghr\ as
\eqn\rone{R^{(1)}_{MN} = \half \bigg( {\partial^2 h^P\,_M\over \partial x^P
\partial x^N} + {\partial^2 h^P\,_N\over \partial x^P \partial x^M} -
{\partial^2 h^P\,_P\over \partial x^M \partial x^N} - {\partial^2 h_{MN}\over
\partial x^P \partial x_P} \bigg),}
where the indices are raised and lowered using the flat Minkowski metric.
Using the above expression for $R^{(1)}_{MN}$, it is easy to calculate the
total energy density for the following metric
\eqn\extremetric{ds^2 = e^{2A(r)} \eta_{\mu\nu} dx^\mu dx^\nu +
                        e^{2B(r)} \delta_{mn} dy^m dy^n,}
as
\eqn\extremtensor{\Theta_{\mu\nu} = {1\over 2\kappa^2} \eta_{\mu\nu} \bigg[
(d -1){\partial^2 e^{2A} \over \partial y^2} + (D - d -1) {\partial^2 e^{2B}
\over \partial y^2} \bigg],}
where $D$ is the space-time dimension; $d - 1$ refers to the spatial dimension
of the black extended object; $\mu,\nu = 0, 1, \cdots, d - 1$;
$ m, n = d, d + 1, \cdots, D - 1$, and $r = \sqrt {\delta_{mn} y^m y^n}$.
Formula \extremtensor\ has been used to calculate the ADM mass per unit
volume for elementary or solitonic (extremal black) $p$-branes in diverse
dimensions in \duflpb, for which the string case ($p = 1$) has been given
before in \dabghr. We would like to stress that eq. \rone\ is not suitable
to be used to calculate the total energy density for a metric such as
\eqn\spherical{ds^2 = - A(r) dt^2 + B(r)dr^2 + r^2 C(r) d\Omega^2_{D - d - 1}
                      + D(r) \delta_{ij} dx^i dx^j,}
which is just the kind to describe a black $(d - 1)$-dimensional extended
object with a spherical symmetry $SO(D - d)$, and where $r > 0$, $i$
runs from $1$ to $d - 1$. The $d \Omega^2_{D -d -1}$ is the line element
on a unit  $(D - d - 1)$-sphere. It is the purpose of this paper to find
the total energy density for the metric \spherical\ by using
eq. \emtensor. Now $g^{(0)}_{MN}$ is:
$g^{(0)}_{00} = - g^{(0)}_{rr} = - g^{(0)}_{ii} = - 1$ and the remaining
metric components are just $r^2$ times those of the unit
$(D - d -1)$-sphere. After a rather tedious calculation, we obtain a very
simple formula for $\Theta_{00}$:
\eqn\sphericaltheta{\eqalign{\Theta_{00} = &- {1\over 2 \kappa^2}
{1\over r^{\tilde d + 1}}\partial_r \bigg[
        (d -1) r^{\tilde d + 1}\partial_r D(r) + (\tilde d + 1)
r^{\tilde d + 1} \partial_r C(r)\cr
&-(\tilde d + 1) r^{\tilde d} \big(B(r)- C(r)\big)\bigg], \cr}}
whose extremal limit goes back to \extremtensor\ as
 $D = A \rightarrow e^{2A}, \quad B = C \rightarrow e^{2B}$, and where
$\tilde d = D - d - 2$.
By using eq. \admass, the corresponding ADM mass per unit volume  is
\eqn\sphericalmass{M_d = - {\Omega_{\tilde d + 1}\over 2 \kappa^2} \bigg[
(\tilde d + 1) r^{\tilde d + 1} \partial_r C(r) + (d - 1) r^{\tilde d + 1}
\partial_r D(r) - (\tilde d + 1) r^{\tilde d} \big(B(r) - C(r)\big)\bigg]_{r
\rightarrow \infty}. }

\newsec{\bf The ADM mass per unit volume for black strings and $p$-branes}

    We would like to calculate the ADM mass per unit volume for the discovered
black strings and $p$-branes in  \refs{\hors, \duflpb}, by using the
 formula developed in the last section. Before jumping
 to a calculation of those ADM masses, we wish to give a brief review of those
black
solutions. As discussed in detail in \refs{\hors, \duflpb}, a
$(D - d - 3)$-brane
solution with magnetic charge ${1\over {\sqrt 2} \kappa} \int F_{d + 1}$ can
be found from the part of the bosonic sector  of the $D$-dimensional
 supergravity
action, which, in terms of canonical variable, is
\eqn\superaction{I_D = {1 \over 2 \kappa^2} \int d^D x \sqrt{- g} \bigg[R -
\half (\partial \phi)^2 - e^{- \alpha (d)\phi} {1\over (d + 1)!}
F^2_{d + 1}\bigg],}
where
\eqn\alphad{\alpha^2 (d) = 4 - {2 d \tilde d \over d + \tilde d},}
with $\tilde d = D - d - 2$. It is found that finding a $(D - d - 3)$-brane
solution from the above action is equivalent to finding a black-hole solution
from the following action
\eqn\holeact{I_{d + 3} = \int d^{d + 3} \hat x \sqrt {- \hat g} \bigg[ \hat R
- \half (\hat \nabla \rho )^2 - \half (\hat \nabla \sigma )^2 - {1\over 2 (d +
 1)!} e^{\beta \rho} {\hat F}^2_{d + 1} \bigg],}
through the following field redefinitions
\eqn\adphi{\eqalign{\beta W &=  - {d(\tilde d - 1)\over (D - 2) (d + 1)}  \rho
 +  \sqrt {\tilde d - 1 \over 2 (D - 2) (d +1)}  \alpha (d) \sigma ,\cr
\beta A &= {d\over D - 2} \rho -  \sqrt { d + 1 \over 2 (D - 2) (\tilde d +1)}
 \alpha (d)   \sigma ,\cr
\beta \phi &= - \alpha (d) \rho - {2d\over \sqrt {2 (D -2)}} {\tilde d -
1\over d + 1} \sigma .\cr}}
The above $\beta$ is,

\eqn\sbeta{\beta = -\sqrt{2 (d + 2)\over d + 1},}
and $W$ and $A$ are defined through
\eqn\metric{ds^2 = e^{2W} d{\hat s}^2 + e^{2A} dx_i dx^i ,}
where $i$ runs from $1$ to $D-d-3$, the spatial dimension of the extended
object, and
$d{\hat s}^2$ is the rescaled metric of the remaining dimension, which is the
one used in the
action \holeact. $A, W$ and $d{\hat s}^2$ are
independent of $x_i$ in order to have $D - d -3$ dimensional translation and
rotation symmetries. The charged  static black-hole  solutions with spherical
symmetry $SO(d + 2)$ to the equations of motion of \holeact\
are asymptotically flat and have a regular horizon. They are
\eqn\blackcon{\eqalign {F &= Q \epsilon_{d+1},\cr
d{\hat s}^2 &= - \bigg[1 - \bigg({r_+\over r}\bigg)^d \bigg]
\bigg[1 - \bigg({r_-\over r}\bigg)^d \bigg]^{1-\gamma d} dt^2 \cr
&\quad + \bigg[1 -  \bigg({r_+\over r}\bigg)^d \bigg]^{- 1}
\bigg[ 1 - \bigg({r_-\over r}\bigg)^d \bigg]^{\gamma - 1} dr^2 \cr
&\quad + r^2 \bigg[1 - \bigg({r_-\over r}\bigg)^d \bigg]^\gamma
d\Omega_{d + 1}^2 ,\cr
e^{\beta \rho} &= \bigg[1 -\bigg({r_-\over r}\bigg)^d \bigg]^{\gamma d}, \cr
\sigma & = 0,\cr}}
where the exponent is
\eqn\expongamma{\gamma = {d + 2\over d (d + 1)}.}
$Q$ is the charge of the black hole, and the $\epsilon_n$ is the volume
element of the unit $n$-sphere. This is a two-parameter family of solutions
labelled by $r_+$ and $r_-$.
These two parameters are algebraically related to the charge and mass of the
black hole. The charge $Q$ is given by
\eqn\charge{ Q = d (r_+ r_-)^{d/2},}
and the mass is proportional to
\eqn\mass{M = - { r_-^d\over d + 1} + r_+^d,}
with a convention-dependent proportionality constant. As discussed in \hors,
if $r_- = 0$, then $F = 0, \rho = 0$ and the above metric reduces to the
$(d + 3)$-dimensional Schwarzschild black hole. At $r = r_+$, the time-like
Killing field becomes null, and there is an event horizon and the curvature is
finite there. Since $\gamma > 0$ for $d \geq 1$, at $r = r_-$, the area of the
sphere goes to zero and there is a curvature singularity. Thus these solutions
describe black holes only as $r_+ > r_-$.

By using \adphi, \metric\ and \blackcon, one obtains black $(D - d -3)$-brane
solutions of \superaction:
\eqn\branecon{\eqalign{F &= Q \epsilon_{d + 1},\cr
ds^2 &= -\bigg[1 - \bigg({r_+\over r}\bigg)^d\bigg] \bigg[1 -
\bigg({r_-\over r}\bigg)^d \bigg]^{\gamma_x - 1} dt^2 \cr
&\quad +\bigg[1 - \bigg({r_+ \over r}\bigg)^d \bigg]^{- 1} \bigg[1 -
\bigg({r_-\over r}\bigg)^d \bigg]^{\gamma_\Omega - 1} dr^2 \cr
&\quad + r^2 \bigg[1 - \bigg({r_-\over r}\bigg)^d \bigg]^{\gamma_\Omega}
d\Omega_{d + 1}^2\cr
&\quad + \bigg[1 - \bigg({r_-\over r}\bigg)^d \bigg]^{\gamma_x} dx_i dx_i ,\cr
e^{-2\phi} &= \bigg[1 - \bigg({r_-\over r}\bigg)^d \bigg]^{\gamma_\phi}, \cr}}
where
\eqn\gammaxop{\eqalign{\gamma_x &= {d\over D - 2}, \cr
\gamma_\Omega &= {\alpha^2 (d)\over 2d},\cr
\gamma_\phi &= \alpha (d). \cr}}
The metric $ds^2$ in \branecon\ fits the general expression \spherical,
but $r > r_-$ rather than $r > 0$. In order to use the ADM mass
formula developed in the last section, we must do the replacement
$r^d \rightarrow r^d + r^d_-$ in \branecon. The resulting metric $ds^2$
looks as
\eqn\rightmetric{\eqalign{ds^2 =& - \bigg[1 + {r^d_+ \over r^d - (r^d_+ -
r^d_-)}\bigg]^{- 1} \bigg[1 + \bigg( {r_-\over r}\bigg)^d \bigg]^{\tilde d
\over D - 2} dt^2 \cr
&+\bigg[1 + \bigg({r_-\over r}\bigg)^d \bigg]^{\tilde d \over D - 2} \bigg[
{1 + {r^d_+ \over r^d - (r^d_+ - r^d_-)}\over 1 + \big({r_- \over r}\big)^d}
dr^2 + r^2 d\Omega^2_{d + 1}\bigg]\cr
&+ \bigg[1 + \bigg({r_-\over r} \bigg)^d\bigg]^{-
{d\over D - 2}} dx^i dx^i. \cr}}
Reading, from the above metric, the  $B(r), C(r)$ and $D(r)$ in \sphericalmass,
 we have the ADM mass per unit volume for the black
$(D - d - 3)$-brane
\eqn\braneadmass{M_{\tilde d} = {\Omega_{d + 1} \over 2 \kappa^2}\big[
(d + 1)r^d_+ - r^d_-\big],}
where $\Omega_n$ is the volume of the  unit $n$-sphere.

\newsec{\bf Discussion}

    From the brief review at the beginning of the last section, we know that
a black $(D - d -3)$-brane solution from the action \superaction\ is
essentially a black-hole solution from the effective action \holeact.
Is there any relation between their ADM masses? Examining eqs. \mass\
and \braneadmass, we find that they are  proportional to each other,
so we expect that they are actually the same, since eq. \mass\ is determined
only up to an overall factor.
It is well known that an ADM mass for a black hole can be read directly
from the asymptotic behaviour of the $00$-component of the metric. One may ask:
 does this rule apply to
black  (non-extremal) $p$-branes with $p > 0$? The answer is simply no.
One can check that only for $\tilde d = 1$, i.e.  a black hole, the ADM
mass from \braneadmass\ agrees with what you read from \rightmetric.
Therefore, finding a black extended object  through a black hole
provides also a simple way to determine its ADM mass per unit volume.

    We have established a Bogolmol'nyi bound for an extremal black, so-called
elementary (electric-like) or solitonic (magnetic-like) $(d - 1)$-brane in
\refs{ \duflfb, \duflpb}, which is
\eqn\ebogbound{\kappa M_d \geq {1\over \sqrt 2} |e_d|,}
for an elementary one, or
\eqn\sbogbound{\kappa M_{\tilde d} \geq {1\over \sqrt 2} |g_{\tilde d}|,}
for a solitonic one, where
\eqn\echarge{ e_d = {1\over {\sqrt 2} \kappa}\int e^{- \alpha(d)}
{^\star F_{d + 1}},}
and
\eqn\scharge{g_{\tilde d} = {1\over {\sqrt 2} \kappa}\int F_{d + 1}.}
We will show that the Bogolmol'nyi bound \sbogbound\ does hold for the black
strings and $p$-branes discussed in the last section. The magnetic charge
for a black $(\tilde d - 1)$-brane, from \charge\ and \branecon, is
\eqn\mcharge{g_{\tilde d} = {\Omega_{d + 1}\over {\sqrt 2}\kappa}
d(r_+ r_-)^{d/2}.}
By noticing that $r_+ \geq r_-$ and
$(d + 1)r^d_+ - r^d_- \geq d r^d_+ \geq d (r_+ r_-)^{d/2}$, we finish our
proof.

    In the previous sections, we have used only the canonical metric, which is
the
one used in the usual Einstein-Hilbert action, and probably the most suitable
 to be used to define the  ADM mass. However, some authors may take,
 for example, the string $\sigma$-model metric to define the ADM mass.
This actually happens in the literature, for example, Giddings and Strominger
used this kind of ADM mass in discussing exact 5-branes in critical superstring
theory \gids. In what follows, an ADM mass per unit $(\tilde d - 1)$-brane
volume calculated in $(n - 1)$-brane metric is given. The relation between
the  $(n - 1)$-brane $\sigma$-model metric and the canonical metric is
\eqn\metrelation{g_{MN} (n) = e^{\alpha (n) \phi \over n}
g_{MN} ({\rm canonical}),}
where $\phi$ is the dilaton field and $\alpha (n)$ is given by \alphad\ upon
taking $d = n$. For example, taking $n = 2$, we have $\alpha (2) = 1$ from
\alphad, eq. \metrelation\ gives just the familiar relation between the string
$\sigma$-model metric and the canonical metric. By using this metric, repeating
what we have done in the last two sections, we find
\eqn\nadmass{M_{\tilde d} = {\Omega_{d + 1} \over 2 \kappa^2}\bigg[
{d (D - 2)\alpha(d) \alpha (n)\over 2n} r^d_- + (d + 1)r^d_+ - r^d_-\bigg].}
Taking $n = 2$ and  $D = 10$ in the above, we have the ADM mass for a black
5-brane:
$M_6 = {3\Omega_3\over 2 \kappa^2}[r^2_+ + r^2_-]$, which is the one used by
Giddings and Strominger in \gids.

\listrefs

\end